\documentclass[11pt]{article}
\usepackage{epsfig}
\usepackage{dcolumn}
\usepackage{bm}
\usepackage{graphicx}
\usepackage{amssymb,amsmath}
\usepackage{multirow}
\usepackage{cite,color,url}
\usepackage{tabu}
\usepackage{array}
\usepackage{here}
\usepackage{physics}
\usepackage[colorlinks=true,urlcolor=blue,anchorcolor=blue
,citecolor=blue,filecolor=blue,linkcolor=blue,menucolor=blue
,linktocpage=true,pdfproducer=medialab,pdfa=true]{hyperref}
\usepackage{colordvi}
\usepackage{epsfig,psfrag,rotating,soul}
\def\beqa{\begin{eqnarray}}
\def\eeqa{\end{eqnarray}}

\evensidemargin \oddsidemargin
\allowdisplaybreaks

\psfrag{lk}{$l_k$}
\psfrag{lm}{$l_m$}
\psfrag{blk}{$\bar {l}_k$}
\psfrag{blm}{$\bar {l}_m$}
\def\thefootnote{\fnsymbol{footnote}}

\let\OLDthebibliography\thebibliography
\renewcommand\thebibliography[1]{
\OLDthebibliography{#1}
\setlength{\parskip}{0pt}
\setlength{\itemsep}{0pt plus 0.3ex}}
\usepackage{makecell}

\begin{document}

\thispagestyle{empty}
\begin{center}
\begin{Large}
\textbf{\textsc{Is the new CDF $M_W$ measurement consistent with the two higgs doublet model?}}
\end{Large}

\vspace{1cm}
{\sc
\vspace{1cm}
{
H. Abouabid$^{1}$%
\footnote{\tt \href{mailto:hamza.abouabid@gmail.com}{hamza.abouabid@gmail.com}}%
, 
A. Arhrib$^{1}$%
\footnote{\tt \href{mailto:aarhrib@gmail.com}{aarhrib@gmail.com}}%
, 
R. Benbrik$^{2}$%
\footnote{\tt
\href{mailto:r.benbrik@uca.ma}{r.benbrik@uca.ma}}
,
M. Krab$^{3}$%
\footnote{\tt
\href{mailto:mohamed.krab@usms.ac.ma}{mohamed.krab@usms.ac.ma}}
,                            
M. Ouchemhou$^{2}$%
\footnote{\tt
\href{mailto:ouchemhou2@gmail.com}{ouchemhou2@gmail.com}}
}

\vspace*{.7cm}

{\sl
$^1$ Abdelmalek Essaadi University, Faculty of Sciences and Techniques,
Tangier, Morocco.}\\
{\sl
$^2$ Laboratory of Fundamental and Applied Physics, Facult\'e Polydisciplinaire de Safi, Sidi Bouzid, BP 4162, Safi, Morocco.}

{\sl
$^3$ Research Laboratory in Physics and Engineering Sciences, Modern and Applied Physics Team,
Polidisciplinary Faculty, Beni Mellal, 23000, Morocco.}
}

\vspace*{.7cm}

\end{center}

\vspace*{0.1cm}

\begin{abstract}
	Motivated by the new CDF measurement of the $W$ boson mass reported recently which clearly illustrates a large deviation compared to the Standard Model (SM) prediction. In the present paper, we study the Two-Higgs Doublet Model (2HDM) contributions to $M_W$ and its phenomenological implications in the case where the heavy CP-even $H$ is identified as the observed Higgs boson with a mass of $125$ GeV. Taking into account theoretical and all the available experimental constraints as well as the new CDF measurement, we demonstrate that the 2HDM parameter space can provide a large correction which predicts the W mass close to the new CDF $M_W$ measurement. It is found 
that $M_{H^\pm}=M_A$  is excluded and the splitting of the charged Higgs boson with all other states is positive. 
We also discuss the consequences on the effective mixing 
angle $\sin^2\theta_{\text{eff}}$ as well as the phenomenological implications on the charged Higgs and CP-odd  Higgs boson decays in 2HDM type I and type X.
\end{abstract}

\newpage
\renewcommand{\thefootnote}{\arabic{footnote}} 
\section{Introduction}
\label{sect:introduction}
ElectroWeak Precision Observables (EWPOs) such as W boson mass, the effective mixing angle $\sin^2\theta_{\text{eff}}$ and the Z boson width  etc, can be used to test the validity Standard Model (SM) and to 
reveal the presence of new physics.\\
After a decade of work, using the data set collected at 8.8 fb$^{-1}$ luminosity and 1.96 TeV center-of mass energy at  the Tevatron, the CDF collaboration  discovered that the W boson has a mass of \cite{CDF,CDF:2021ydv}: 
\begin{eqnarray}
M_W^{\text{CDF}}=80.4435\pm 0.0094 \ \rm{GeV}. 
\label{MWCDF}
\end{eqnarray}
The precision with which this measurement was carried out, 0.01\%,  exceeds all previous measurements combined.
In addition, the new value agrees with many previous W mass measurements, 
but there are also some disagreements \cite{ParticleDataGroup:2020ssz}. 
Therefore, future measurements will be needed to further shed light on the outcome.
The above measurement  should be compared to the SM prediction \cite{ParticleDataGroup:2020ssz,Awramik:2003rn},
\begin{eqnarray}
M_W^{\text{SM}}= 80.357 \pm 0.006\ \ \rm{GeV}.
\label{MWSM}
\end{eqnarray}
Note that the above value is based on complex SM radiative corrections that closely relate the mass of the W to the measurements 
of the masses of the top quark and the Higgs boson.
It is clear that $M_W^{\text{CDF}}$ presents a deviation from $M_W^{\text{SM}}$ with a significance of 7.0 $\sigma$.

In the past, during the LEP era, it was well known that the global fit of the SM to LEP and SLC data has been used to predict 
the existence of heavy top quark  and relatively light Higgs boson well before their discovery at Tevatron and LHC respectively.
Although the CDF measurement needs to be confirmed shortly, 
it is quite likely that the difference between the experimental value and the expected SM value is due to a 
non-decoupled new particle or a new interaction. If it is the case, there is a chance that these new phenomena 
will show up in future experiments.

Moreover, it is well known that the discrepancy of $M_W^{\text{CDF}}$ 
from the SM prediction can be parameterized in terms of the oblique parameters, $S$, $T$ and $U$  
\cite{Peskin:1990zt, Peskin:1991sw} which are a combination of gauge bosons self energies. All
new particles, if not too heavy and interact with the photon, $W$ and $Z$ bosons, will contribute to $S$, $T$ and $U$ and can therefore 
reduce the tension between $M_W^{\text{CDF}}$ and $M_W^{\text{SM}}$.

In the present paper, we will discuss the implications of the new CDF measurement on the Two-Higgs 
Doublet Model (2HDM) which predicts in its spectrum 2 CP-even, $h$ and $H$ (with $M_h < M_H$), one CP-odd $A$ and a pair of charged Higgs $H^\pm$. Recently, there have been several studies addressing a similar issue 
within: the 2HDM
\cite{Broggio:2014mna,Lu:2022bgw, Fan:2022dck, Song:2022xts, Bahl:2022xzi, Babu:2022pdn, Biekotter:2022abc, Han:2022juu, Heo:2022dey, Ahn:2022xeq, Benbrik:2022dja,Arcadi:2022dmt, Ghorbani:2022vtv, Lee:2022gyf}, triplet extension \cite{Du:2022brr,Ghoshal:2022vzo,Kanemura:2022ahw, Addazi:2022fbj}
 and also other SM extensions \cite{Yang:2022gvz, Strumia:2022qkt, Sakurai:2022hwh, Liu:2022jdq, DiLuzio:2022xns, Asadi:2022xiy, Heckman:2022the, Bagnaschi:2022whn, Paul:2022dds, Balkin:2022glu, Endo:2022kiw, Zheng:2022irz, Carpenter:2022oyg, Popov:2022ldh, Chowdhury:2022moc,Cirigliano:2022qdm,Bhaskar:2022vgk,Baek:2022agi,Cao:2022mif,Kawamura:2022uft, Nagao:2022oin, Zhang:2022nnh, Borah:2022zim, Cheng:2022aau, Batra:2022org}.
In this study, we identify the observed SM Higgs with $H$ whose properties are consistent with the LHC measurements and assume that the second CP-even Higgs is lighter than $125$ GeV.
We will explain how the 2HDM can solve the tension between CDF measurement and the SM prediction 
and give some phenomenological implications on charged Higgs and CP-odd Higgs boson decays both in 2HDM type I and type X.

The paper is organized as follows, in the second section we briefly introduce the set-up of the 2HDM and 
give the $S$, $T$ and $U$ formalism for the computation of $M_W^{\text{2HDM}}$ and $\sin^2\theta_{\text{eff}}^{\text{2HDM}}$.
In the third section, we present the details of our scan as well as the theoretical and experimental constraints used 
to constrain the parameter space. We then present our main result and explain how the 2HDM spectrum can 
predict the W mass that is close to the new CDF measurement. In addition, we give phenomenological implications for the charged Higgs and CP-odd boson decays within the allowed parameter space. We conclude in section four.

\section{$M_W$ in the 2HDM} 
\label{sect:2HDM-review}
\setcounter{equation}{0}

The 2HDM framework is one of the simplest extensions of the SM Higgs sector. 
It contains two Higgs doublet fields, $\phi_{1}$ and $\phi_2$, that can interact with fermions and 
gauge bosons to generate their masses. \\
The most general scalar potential which is invariant under $SU_L(2)\times U_Y(1)$ and CP-conserving 
can be written as
\begin{eqnarray}
V(\phi_1,\phi_2) &=& m_{11}^2(\phi_1^\dagger\phi_1) +
m_{22}^2(\phi_2^\dagger\phi_2) -
[ m_{12}^2(\phi_1^\dagger\phi_2)+\text{h.c.}] \nonumber\\ 
&+& \frac12\lambda_1(\phi_1^\dagger\phi_1)^2 +
\frac12\lambda_2(\phi_2^\dagger\phi_2)^2 +
\lambda_3(\phi_1^\dagger\phi_1)(\phi_2^\dagger\phi_2)~\nonumber\\
&+&\lambda_4(\phi_1^\dagger\phi_2)(\phi_2^\dagger\phi_1)
+\frac12\left[\lambda_5(\phi_1^\dagger\phi_2)^2 +\rm{h.c.}\right],
\label{CTHDMpot}
\end{eqnarray}
where $\lambda_{1,2,3,4,5}$ as well as $m_{11}^2$ and $m_{22}^2$ are chosen to be real. 
If both Higgs fields interact with all SM fermions, like in the SM, 
we end up with Flavour Changing Neutral Currents (FCNCs) 
at the tree-level in the Yukawa sector. In order to avoid such FCNCs, a discrete $Z_2$ symmetry is
introduced to prevent large tree-level FCNCs \cite{Glashow:1976nt,Branco:2011iw}. 
Such a discrete symmetry is imposed both on the Yukawa sector as well as the scalar potential where we allow for a  soft violation of $Z_2$  by $ m_{12}^2(\phi_1^\dagger\phi_2)$ term. Moreover, under the $Z_2$ symmetry, there are four possible types of Yukawa sector: type I, type II, type X (or lepton-specific) and type Y (or flipped). Here, in this work, we shall focus on type I and X models. In the 2HDM type-I, $\phi_2$ doublet couples to all the SM fermions exactly like in the SM 
while in the 2HDM type-X all the quarks couple to $\phi_2$   and the charged leptons couple to $\phi_1$.

Using the two minimization conditions, $m^2_{11}$ and $m^2_{22}$ can be expressed as functions of other parameters.
The combination of $v_1$ and $v_2$ is fixed from the electroweak scale: $v^2_1 + v^2_2 = v^2 \simeq (246 ~\text{GeV})^2$. We thus end up with 7 independent parameters, namely $\lambda_{1,2,3,4,5}$, $m^2_{12}$ and $\tan\beta$. Alternatively, the set $M_h$, $M_H$, $M_A$, $M_{H^\pm}$, $\sin(\beta-\alpha)$, $\tan\beta$ and $m^2_{12}$ can be chosen instead. The angle $\alpha$ is the mixing angle between the two CP-even scalars $h$ and $H$, while $\beta$ is defined as the ratio of the vevs, $\tan\beta = v_2/v_1$.  

The 2HDM contribution to the EWPOs can be described by the oblique parameters formalism, which is a good 
one for  new physics. A convenient parametrization  of such formalism is given by the well known parameters $S$, $T$ and $U$ \cite{Peskin:1990zt, Peskin:1991sw}. In the 2HDM, the $\rho$ parameter, which is the ratio of neutral and charged  current at small momentum transfert, is related to the oblique parameter $T$. Such a contribution is controlled by the so-called custodial symmetry  to preserve the tree-level value of $\rho$ parameter, $\rho =M_W^2/(c_W^2  M_Z^2)\approx  1$, which is in good agreement with experiments. 
As discussed in the literature, in the SM the custodial symmetry is broken both by the Hypercharge as well as by the different sizes of the Yukawas, while in the 2HDM the custodial symmetry can be restored in the scalar sector so long the Higgs states are degenerate in mass.         

In general, the contribution of 2HDM to $W$ boson mass can be expressed in terms of the parameters $S$, $T$ and $U$ \cite{Peskin:1991sw}, i.e. 

\begin{equation}
\Delta M^2_W = \frac{\alpha_0 c^2_W M^2_Z}{c^2_W - s^2_W} \left[-\frac{1}{2} S + c^2_W T + \frac{c^2_W - s^2_W}{4 s^2_W} U \right], \label{Mw_2HDM}
\end{equation}
where $\Delta M^2_W=(M_W^{\text{2HDM}})^2 - (M_W^{\text{SM}})^2$, $M_Z$ is the Z boson mass , $c_W=M_W^{\text{SM}}/M_Z$ and $s_W$ are respectively cosine and sine of the weak mixing angle ($s_W^2=1-c_W^2$) and $\alpha_0$ is the fine structure constant at the Thomson limit. 

We also study the effects of the 2HDM spectrum on the effective weak  mixing angle, $\sin^2\theta_{\text{eff}}$.
This is computed using the following relation \cite{Peskin:1991sw}:
\begin{equation}
\Delta\sin^2\theta_{\text{eff}} = \frac{\alpha_0}{c^2_W - s^2_W} \left[\frac{1}{4} S - s^2_W c^2_W T \right].  \label{sin2theta} 
\end{equation} 
Where $\Delta\sin^2\theta_{\text{eff}}$ is the difference between the 2HDM and the SM value.
Note that in both Eq. (\ref{Mw_2HDM}) and Eq. (\ref{sin2theta}), the dominant correction to the $W$ boson mass and $\sin^2\theta_{\text{eff}}$ comes from $T (\equiv\delta\rho/\alpha_0)$ parameter, which is sensitive to the mass splitting of 2HDM scalar particles. The size of $T$ parameter could be viewed as the amount of violation of the Custodial symmetry by the 2HDM spectrum. The SM values used in our calculation are given in Ref. \cite{ParticleDataGroup:2020ssz}.  
Analytic expressions for $S$, $T$ and $U$ parameters in the 2HDM are given in \cite{Eriksson:2009ws}. 


\section{Results and discussions}
\label{sect:results}
To study the implication of  the new  CDF measurement on the 2HDM, we consider the 
2HDM type I\footnote{The 2HDM contribution to $\Delta M_W$ is expected to be the same in all Yukawa types at the one-loop level. The main difference comes from LHC constraints on Higgs physics.}  and perform a systematic scan over its parameter space. The scan is done using the public program \texttt{2HDMC-1.8.0} \cite{Eriksson:2009ws}. 
We assume that the CP-even Higgs boson $H$ is the observed SM-like Higgs with 
$M_H = 125.09$ GeV \cite{ATLAS:2015yey}  whose properties are consistent with the LHC measurements.
In addition, we suppose that the light CP-even $h$ is lighter than 125 GeV and check that it is consistent with the 
previous negative LEP and LHC searches.
We randomly sample the remaining model parameters within the following ranges:      
\begin{eqnarray}
&&M_h = 15 ~\text{--}~ 120~\text{GeV}; ~~M_A = 15~\text{--}~700~\text{GeV};\\\nonumber
&& M_{H^\pm} = 80~\text{--}~700~\text{GeV}; ~~\sin(\beta-\alpha) = -0.5~\text{--}~0.5; \\\nonumber
&& \tan\beta = 2~\text{--}~25; ~~m_{12}^2 = 0~\text{--}~M_{h}^2\sin\beta\cos\beta. 
\end{eqnarray}

During the scan, the following theoretical and experimental constraints are fulfilled 
\begin{itemize}
	\item Unitarity, perturbativity and vaccum stability are imposed via \texttt{2HDMC}.  
	
	\item Exclusion bounds at $95\%$ Confidence Level (CL) from additional Higgs bosons are enforced via \texttt{HiggsBounds-5.9.0} \cite{Bechtle:2020pkv}. 
	
	\item Compliance with  SM-like  Higgs state measurements are enforced via \texttt{HiggsSignals-2.6.0} \cite{Bechtle:2020uwn}.
	
	\item Constraints from flavor physics are enforced using the result given in Ref. \cite{Haller:2018nnx}. Related observables are calculated using the program \texttt{SuperIso v4.1} \cite{Mahmoudi:2008tp}.
	
	\item Compatibility with the $Z$ width measurement from LEP, $\Gamma_Z = 2.4952 \pm 0.0023$ GeV \cite{ALEPH:2005ab}. The partial width $\Gamma(Z \rightarrow h A)$, when is kinematically open, was chosen to satisfy the 
	$2\sigma$ experimental uncertainty of the measurement.
\end{itemize}

Once we get the allowed parameter space that satisfies all the above theoretical and experimental constraints, 
we then apply the following  $\chi^2_{M^{\text{CDF}}_W}$ test where we take only 
points that are within  $2\sigma$ of the new CDF measurement.

\begin{equation}
\chi^2_{M^{\text{CDF}}_W} = \frac{(M^{\text{2HDM}}_W - M^{\text{CDF}}_W)^2}{(\Delta M^{\text{CDF}}_W)^2},
\end{equation}   
where $\Delta M^{\text{CDF}}_W = 0.0094$ GeV is the uncertainty of the new CDF measurement (see Eq. \ref{MWCDF}).

\begin{figure}[H]
\centering
\includegraphics[scale=0.49]{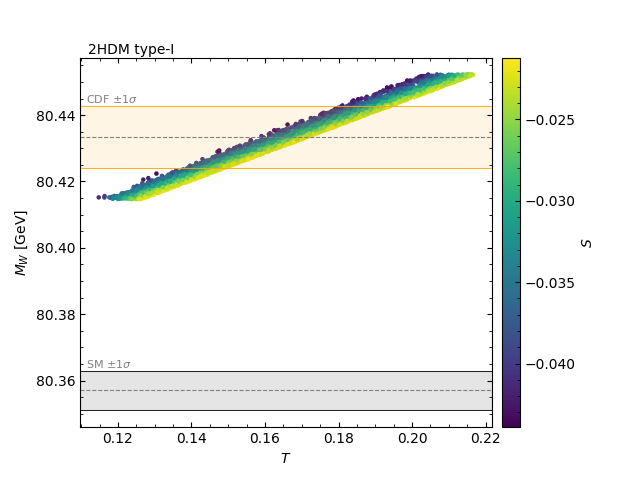}
\includegraphics[scale=0.487]{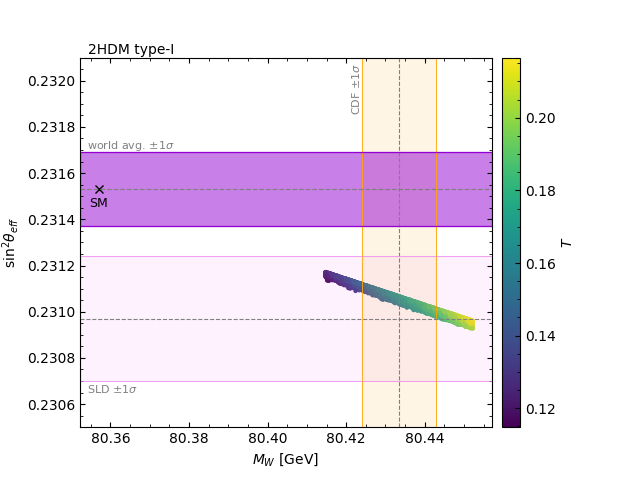}
\caption{Left: The 2HDM prediction for the $W$ boson mass as a function of $T$, with the color bar showing the size of $S$. The light orange band indicates the new CDF measurement and the associated $1~\sigma$ uncertainty. The SM prediction for $M_W$ is given by the light black region within $\pm1\sigma$.
Right: The 2HDM prediction for $M_W$ and $\sin^2\theta_{\text{eff}}$, with the color bar refers to $T$. The light orange band indicates the new CDF measurement and the associated $1~\sigma$ uncertainty. The light and dark violet regions represent the SLD and world average results for $\sin^2\theta_{\text{eff}}$ within the $1\sigma$ level, respectively. The SM prediction for $M_W$ and $\sin^2\theta_{\text{eff}}$ is given by the black cross.}
\label{fig_Mw}
\end{figure}

In Figure \ref{fig_Mw} (left panel), 
we present the 2HDM prediction for the W boson mass in the allowed parameter 
as a function of $T$, where the color map shows the possible size of $S$. The light orange band shows the new CDF result for  $M_W$  within the $1\sigma$ uncertainty. We also depict via light black region the SM prediction at the $1\sigma$ level. As expected from Eq. (\ref{Mw_2HDM}), the dominant contribution arises from the $T$ parameter which is almost a linear relation. $M_W$ receives a negative correction from $S$ as indicated by the color code. The dependence of $U$ to $M_W$ is found to be negligible compared to $T$. It is obvious from Eq. (\ref{Mw_2HDM})
that negative range of the $S$ parameter and positive values of $T$ and $U$ are indeed favored by the new CDF  measurement for $M_W$.  Therefore, close to the alignment limit $\cos(\beta-\alpha)\approx 1$, the degenerate case $M_A=M_{H^\pm}$ of the 2HDM  is excluded by this new measurement because it would make $T$ vanish.
We note that the allowed range for $S$ and $T$ parameters is consistent with the recent results found in the literature 
\cite{Lu:2022bgw,Strumia:2022qkt,deBlas:2022hdk,DiLuzio:2022xns}.


Another important precision observable is the effective weak mixing angle, $\sin^2\theta_{\text{eff}}$. The 2HDM prediction for $M_W$ and $\sin^2\theta_{\text{eff}}$ for the allowed points is depicted in the right panel of Figure \ref{fig_Mw}. The light violet band indicates the SLD $\sin^2\theta_{\text{eff}}$ measurement within the $1\sigma$ uncertainty \cite{ALEPH:2005ab}. We also show via the dark violet region the world average value for $\sin^2\theta_{\text{eff}}$ at $1\sigma$ level (for the purpose of comparison) \cite{ALEPH:2005ab}. It can be seen clearly  that $M_W$ within the $2\sigma$ level is in good compliance with the SLD measurement which is not the case for the world average value.

\begin{figure}[H]
	\centering
	\includegraphics[scale=0.5]{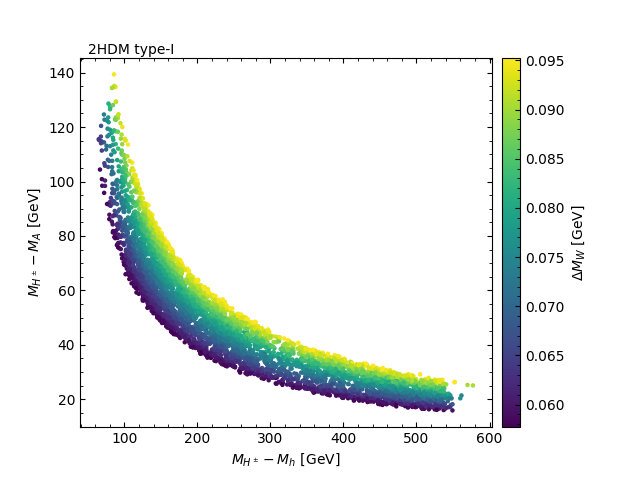}
	\includegraphics[scale=0.5]{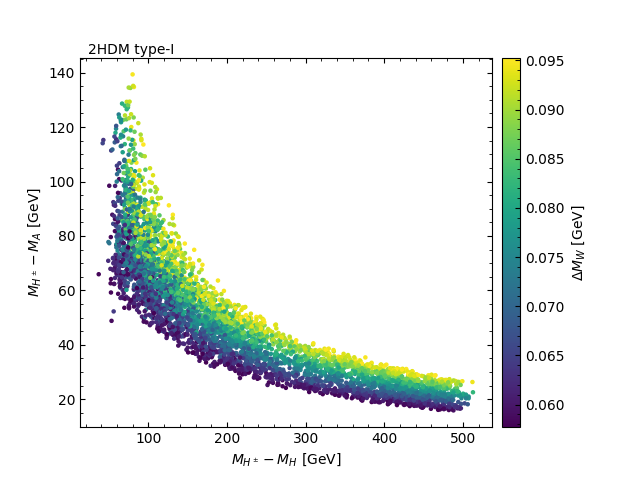}
\caption{Points from the  scan in the ($M_{H^\pm}-M_h, M_{H^\pm}-M_A$) plane (left) and 
($M_{H^\pm}-M_H, M_{H^\pm}-M_A$) plane (right) planes in 2HDM type I. 
The color code indicates the shift from the SM prediction for $M_W$.} \label{splitting}
\end{figure}

As a first implication, we investigate the impact of the new CDF measurement on the spectrum of the 2HDM with inverted hierarchy. In the 2HDM with normal hierarchy: $M_h=125$ GeV and $m_H>m_h$, it was shown recently in literature that both $M_{H^\pm} > M_{H}, M_A$ and $M_{H^\pm} < M_{H}, M_A$ are favored by the new CDF measurement for $M_W$, 
 whereas the case where $M_{H^\pm} \sim M_H \sim M_A$ is disfavored in the alignment limit  of the 2HDM \cite{Bahl:2022xzi,Ahn:2022xeq}. \\
In the 2HDM with inverted hierarchy, we illustrate in Figure \ref{splitting}  the splitting $M_{H^\pm}-M_A$ as a function of $M_{H^\pm}-M_h$ (left panel) and as a function of $M_{H^\pm}-M_H$ (right panel) with $\Delta M_W$ color coded and 
represented in the vertical panel. It is clear that the CDF $M_W$ measurement force $M_{H^\pm}$ to be always heavier than the neutral Higgses. The  cases $M_{H^\pm}<M_h, M_H, M_A$ are excluded by the fact that it produces a negative or small $T$ and therefore can not account for the new CDF measurement.
Another outcome of the scan is that the new CDF $M_W$ measurement push the charged Higgs mass to be 
larger than 161 GeV. 
For completeness, we show in Figure \ref{Mw_appendix} of Appendix A the full scan for $M_W$ as a function of $S$ (left plot) and  as a function of $T$ (right plot). The green points are the one that reproduce the new CDF measurement while the blue points does not give the correct $M_W$ mass. Similarly, we illustrate in Figure \ref{splitting_appendix} of Appendix A the full scan for $M_{H^\pm}-M_A$ as a function of $M_{H^\pm}-M_h$
(left plot) and  $M_{H^\pm}-M_A$ as a function of $M_{H^\pm}-M_H$ (right plot). Only the green band 
reproduce the new CDF measurement. It is clear from the plots that the new CDF measurement 
push the charged Higgs mass to be larger than 161 GeV.

The results shown in the previous plots are for 2HDM type I. For 2HDM type X, we obtain similar plots.
The raison is that $\Delta M_W^{\text{2HDM}}$ depend only on $S$, $T$ and $U$ which are a combination of gauge boson self energies that contains the contribution of the additional Higgs bosons. 
The interaction of the gauge bosons to the Higgs boson does not depend on the Yukawa type. The only difference between type I and type X would come from the LHC constraints. Such constraints depend on the production cross section of the Higgs and its branching fractions which are sensitive to the Yukawa type. 
Note also that the combination of EWPOs  as well as the theoretical constraints set a limit on the masses of the heavy states 
$H^\pm$ and $A^0$ to be less than about 600 GeV.

 
\begin{figure}
	\centering
	\includegraphics[scale=0.5]{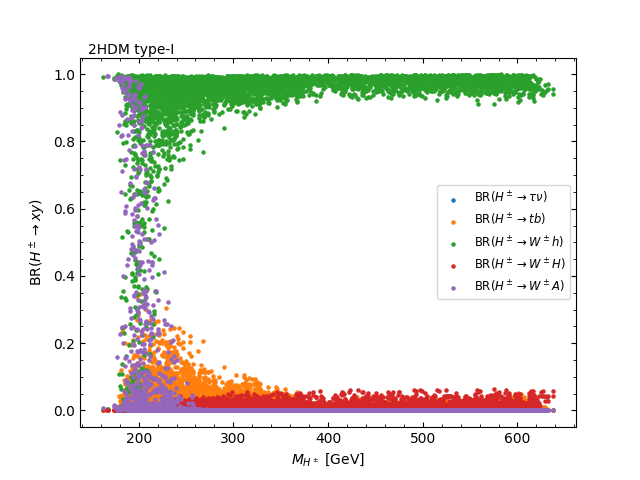} 
	\includegraphics[scale=0.5]{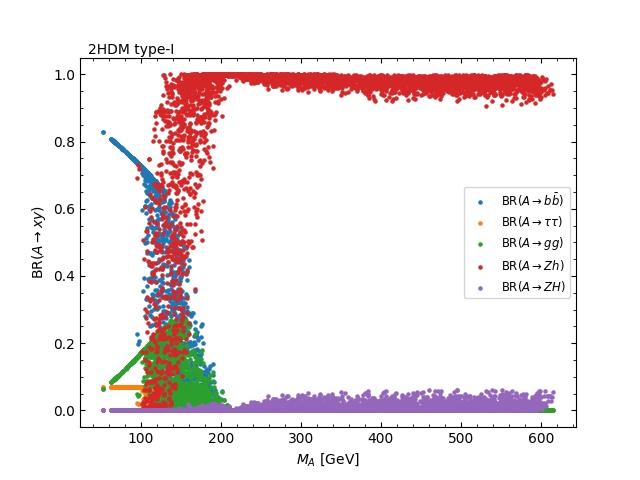}
	\includegraphics[scale=0.5]{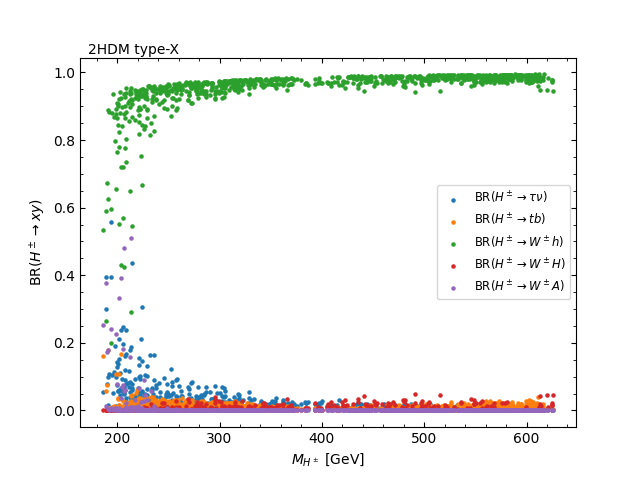} 
	\includegraphics[scale=0.5]{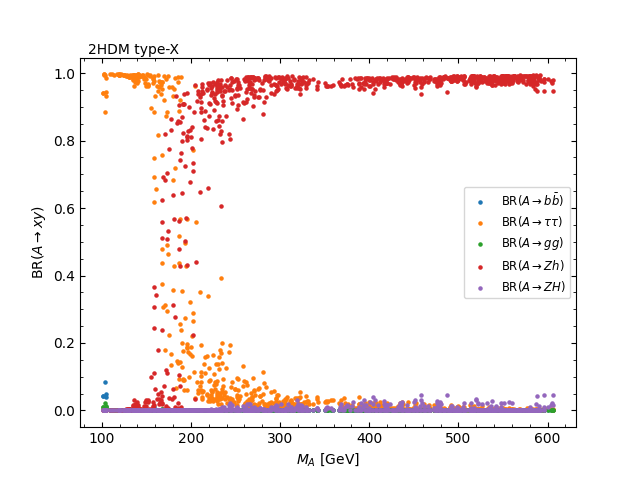}
\caption{Branching ratios of the charged Higgs boson as a function of $M_{H^\pm}$ 
(left) and CP-odd Higgs boson as a function of $M_A$ (right). The upper panels are for 2HDM type I while the lower panels are for 2HDM type X.} 
	\label{BRs_HpA_fig}
\end{figure}

We move now to discuss phenomenological consequences on the charged Higgs, CP-odd Higgs and the light CP-even  Higgs decays in 2HDM type I and type X.
In the upper left-upper panel of Figure \ref{BRs_HpA_fig}, we depict the branching ratios of $H^\pm$ 
as a function of $M_{H^\pm}$. As one can see, the dominant decay 
of the charged Higgs are the bosonic channels: $H^\pm \rightarrow W^\pm h$ and $H^\pm \rightarrow W^\pm A$. 
For charged Higgs mass less than 200 GeV, both channels $H^\pm \rightarrow W^\pm h$ and $H^\pm \rightarrow W^\pm A$
compete. The decay  $H^\pm \rightarrow W^\pm h$ enjoy more phase space because $M_h<125$ GeV, while the decay 
$H^\pm \rightarrow W^\pm A$  is open only for a small portion of the phase space when $M_{H^\pm}-M_A>80$ GeV.
In fact, when $H^\pm \rightarrow W^\pm A$ is open it compete strongly with  $H^\pm \rightarrow W^\pm h$
because the coupling $H^\pm W^\pm A$ is a pure gauge coupling while  $H^\pm W^\pm h$ is suppressed by $\cos(\beta-\alpha)$. This is why we can see that, in some cases, the channel $H^\pm \rightarrow W^\pm A$ is the dominant one.
The channel $H^\pm \to W^\pm H$ is negligible since it is suppressed 
by the parameter $\sin(\beta-\alpha)$ which is small in our scenario.
Note that in 2HDM type I, charged Higgs coupling to fermions is proportional to $1/\tan\beta$ which makes 
$H^\pm\to tb , \tau \nu$ channels rather suppressed. We stress here that the dominance of the bosonic channels 
has been discussed previously in  \cite{Arhrib:2016wpw,Bahl:2021str,Arhrib:2021xmc,Wang:2021pxc,Arhrib:2021yqf}.

In Figure \ref{BRs_HpA_fig} (upper right panel) we depict the branching ratios of $A$ 
as a function of $M_A$. It is clearly visible that before $Zh$ threshold, the pseudo-scalar boson $A$ decays dominantly into a pair of bottom quarks followed by $A \rightarrow gg$ and $A \rightarrow \tau\tau$ decays. The loop decay $A\to \gamma\gamma$ is suppressed by $1/\tan^2\beta$ and is smaller than $10^{-3}$.
Once we cross $Zh$ threshold,  the decay channel 
$A \rightarrow Z h$  becomes the dominant one when $M_A>M_Z+M_h$ since the coupling $AZh$ is proportional to 
$\cos(\beta-\alpha)$ which is close to unity in our scenario. The channel $A\to ZH$ is suppressed by  $\sin(\beta-\alpha)$ 
being close to vanish. Note that $A \rightarrow H^\pm W^\mp$ mode is kinematically not open after taking into account the $M_W$ CDF measurement since $M_{A} < M_{H^\pm}$. This is, indeed, a strong effect of the new CDF measurement  which closes the possibility to probe the charged Higgs boson via $A \rightarrow H^\pm W^\mp$ and/or $H \rightarrow H^\pm W^\mp$ decay channels.    

Note that in the case of 2HDM type X as illustrated in 
Figure \ref{BRs_HpA_fig} (lower  panels), we found that the charged Higgs decay dominantly to $W^+h$  once $W^+h$ threshold is crossed. Before $W^+h$ threshold, there is a strong competition between $\tau\nu$, $W^+h$  and $W^+A$ channels, see Figure \ref{BRs_HpA_fig} (lower and upper left panels).
Similarly for the decay of the CP-odd Higgs. Before the opening of $A\to Zh$, the channel $A\to\tau \tau $ is the dominant one
and gets suppressed once $A\to Zh$ is open, see Figure \ref{BRs_HpA_fig} (lower right panel). 

\begin{figure}[H]
	\centering
	\includegraphics[scale=0.5]{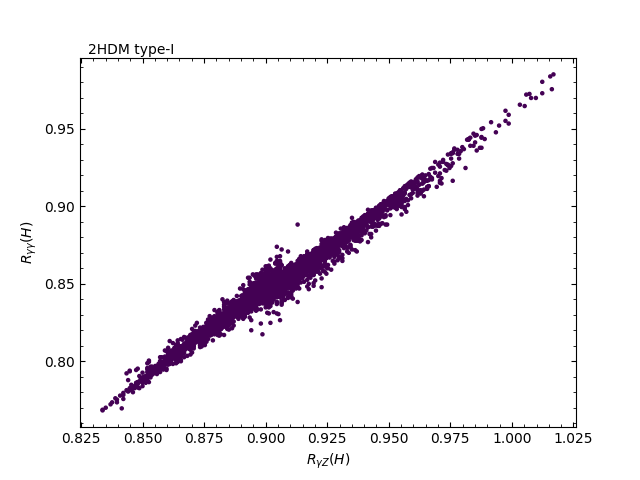} \includegraphics[scale=0.5]{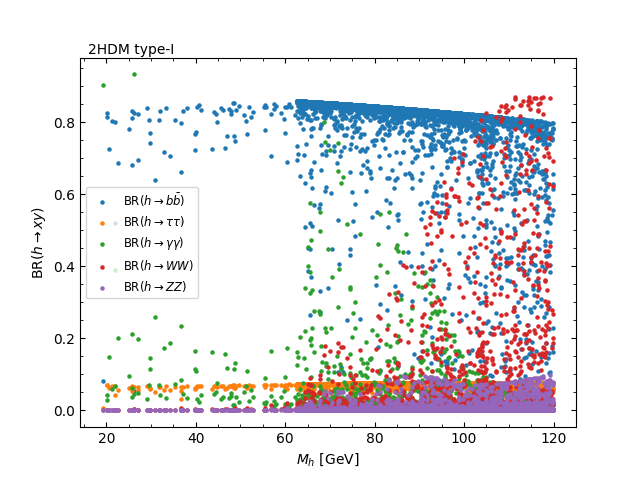}
	\includegraphics[scale=0.5]{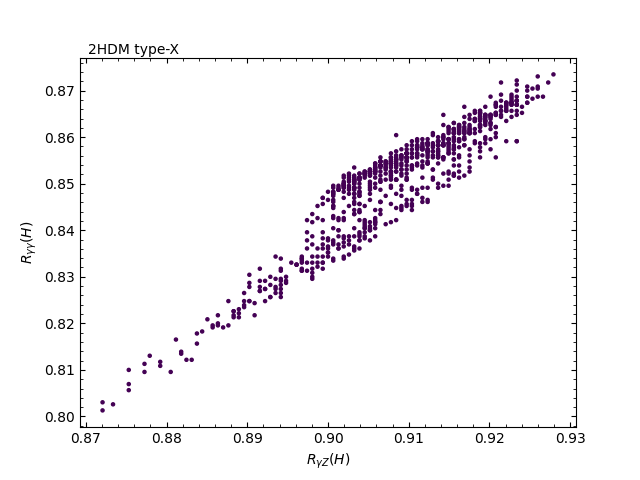} \includegraphics[scale=0.5]{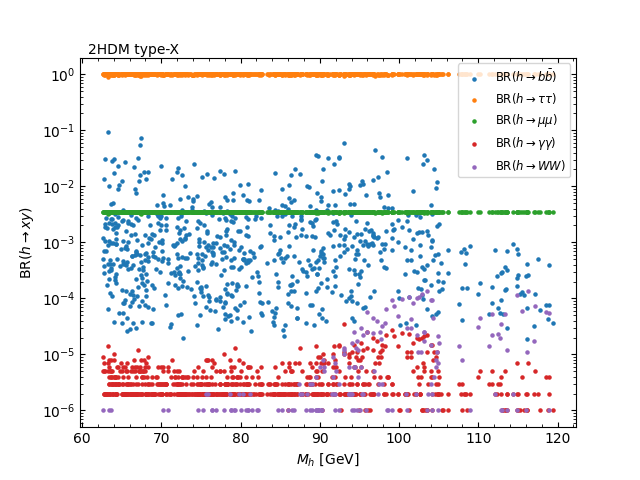}
	\caption{Left: Correlation between  $R_{\gamma\gamma}(H)$
		and $R_{\gamma Z}(H)$  for the SM-like Higgs. Right: Branching fractions for the light CP even. The plots are for 2HDM type I (upper panels) and type X (lower panels).} 
	\label{H-gaga-gaz}
\end{figure}

Before we end this section, we illustrate in Figure \ref{H-gaga-gaz} (left upper and lower panels) the correlation between 
$R_{\gamma\gamma}(H)= \text{BR}(H \rightarrow \gamma\gamma)/\text{BR}(H \rightarrow \gamma\gamma)_{\text{SM}}$
and $R_{\gamma Z}(H)=\text{BR}(H \rightarrow \gamma Z)/\text{BR}(H \rightarrow \gamma Z)_{\text{SM}}$ for the SM-like Higgs both in 2HDM type I and type X. It is clear that $R_{\gamma Z}(H)$ is all time in the range $[0.85, 1]$
while  $R_{\gamma \gamma}(H) \in [0.77, 0.97]$. One can see that they are linearly correlated with 
$R_{\gamma Z}(H)$ slightly larger than $R_{\gamma \gamma}(H)$ both  in 2HDM type I and type X. However, in 2HDM type X, the ranges of  $R_{\gamma \gamma}(H)$ and $R_{\gamma Z}(H) $ are a bit smaller compared to 2HDM type I.  
In Figure \ref{H-gaga-gaz} (upper right panel) we illustrate the branching fractions of the light CP-even $h$ in 2HDM type I. 
One can read that before the $WW^*$ threshold, the dominant decay of $h$ is into a pair of bottom followed 
by $h\to \gamma \gamma$ which could reach 90\% close to the fermiophobic limit. In 2HDM type X 
as illustrated in Figure \ref{H-gaga-gaz} (lower right panel),  the decay $h \rightarrow \tau\tau$ is the dominant decay channel with a branching ratio almost above 99\%. One can also read that 
BR$(h \rightarrow b\bar{b})$ does not exceed  10\% and  BR$(h \rightarrow \mu^+\mu^-)$  becomes of the order 
$3\times 10^{-3}$. \\
It is also clear from Figure \ref{H-gaga-gaz} that in 2HDM type I one can have the CP-even $h$ as light as 20 GeV
while in 2HDM type X we have $m_h>63$ GeV. 
 
\section{Conclusion}

\label{sect:conclusion}
Recently, CDF release its new measurement for W boson mass with unprecedented accuracy.
The new CDF measurement presents a deviation from the SM prediction with a significance of 7.0 $\sigma$.
We have shown that in 2HDM with an inverted hierarchy, it is possible to solve the tension between the new CDF $M_W$ measurement and the SM prediction.  We found that to comply with the CDF measurement we need a positive T 
and this is possible in the case where $M_{H^\pm}>M_h, M_H, M_A$. The case of $M_{H^\pm}<M_h, M_H, M_A$ 
fail to reproduce the correct $M_W$ measurement. Note also that the new CDF measurement for $M_W$
push the charged Higgs to be larger than about 161 GeV.  It is also found  that the degenerate case 
$M_{H^\pm}=M_A$ which leads to a very small T parameter being excluded.
We have presented a phenomenology of charged Higgs, CP-odd and the light CP-even higgsses by 
illustrating their branching fractions both in 2HDM type I and type X. In the case of charged Higgs, 
we observe that the bosonic decay
$H^\pm \to W^\pm h$ is the dominant one both for 2HDM type I and type X. While for the case of CP-odd, 
 we have noticed that  $A\to Zh$, once open,  is the dominant mode both for 2HDM type I and X. 
 However, before $Zh$ threshold,  $A\to bb$ 
is the dominant mode for 2HDM type I while $A\to \tau \tau$ would dominate in the case of 2HDM type X.
We have shown also that in this scenario and within 2HDM type X, BR$(h \to \mu^+\mu^-)$ could be of the order 
$3\times 10^{-3}$.

{\bf{Note Added: }} While we were finishing this work, we received a paper \cite{Lee:2022gyf} dealing with similar 2HDM study both in normal and inverted hierarchy. In the case of inverted hierarchy our results are in good agreement. 

{\bf{ Acknowledgements:}} This work  is supported by the Moroccan Ministry of Higher Education and Scientific Research MESRSFC and CNRST: Projet PPR/2015/6.



%
%

\newpage

\newpage
\appendix
\renewcommand{\thefigure}{A\arabic{figure}}
\setcounter{figure}{0}
\section*{Appendix A} \label{appendix}
\begin{figure}[H]
	\centering
	\includegraphics[scale=0.45]{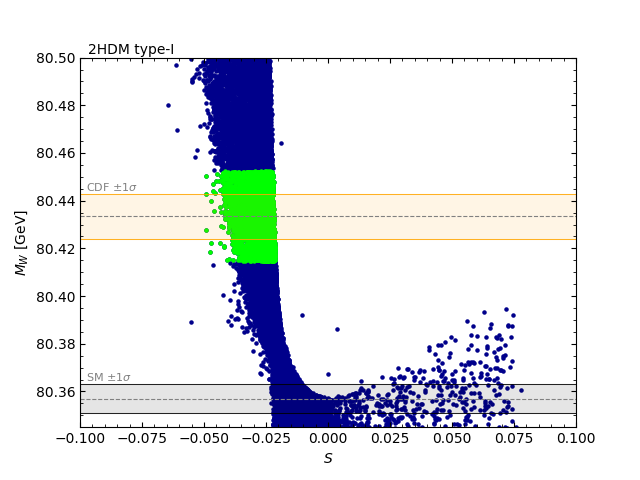} 
	\includegraphics[scale=0.45]{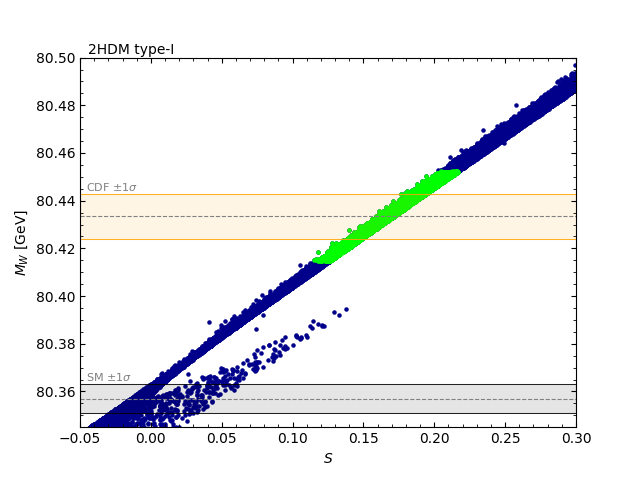}
	\caption{The 2HDM prediction for the W boson mass as a function of $S$ (left panel) and  $T$ (right panel). The light green band represents points within the $2\sigma$ CDF $M_W$ measurement.} \label{Mw_appendix}
\end{figure}
\begin{figure}[H]
	\centering
	\includegraphics[scale=0.45]{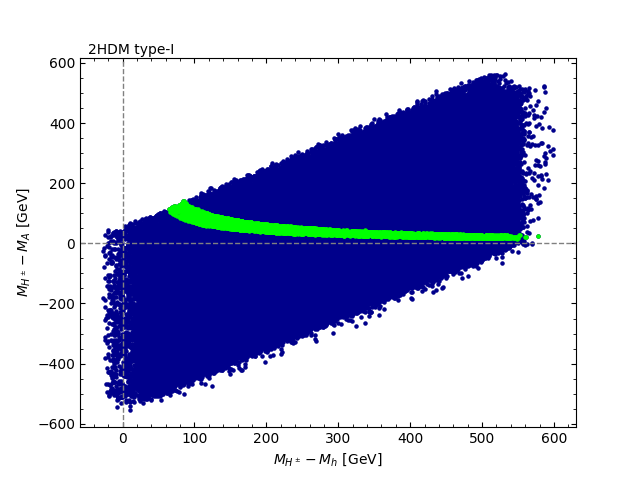} 
	\includegraphics[scale=0.45]{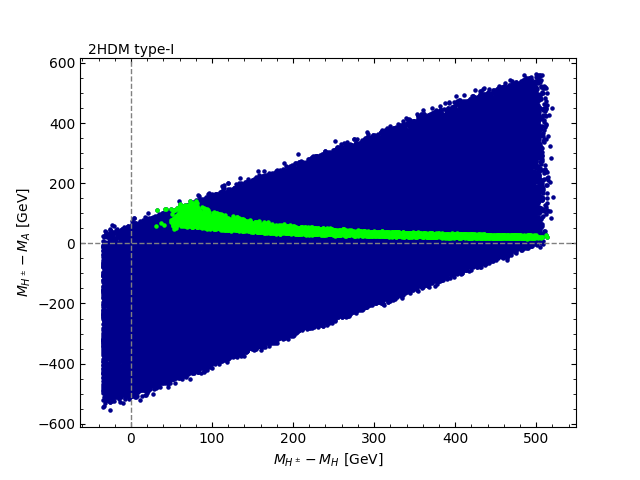}
\caption{Points from the scan in the ($M_{H^\pm}-M_h, M_{H^\pm}-M_A$) plane (left panel) and 
 ($M_{H^\pm}-M_H, M_{H^\pm}-M_A$) plane (right panel). 
 The light green band represents points within the $2\sigma$ CDF $M_W$ measurement.} 
 \label{splitting_appendix}
\end{figure}


\begin{thebibliography}{100}
	
	\bibitem{CDF} T. Aaltonen et al. (CDF collaboration), Science 376, 170-176 (2022). 
	
	\bibitem{CDF:2021ydv}
	T.~Aaltonen \textit{et al.} [CDF],
	[arXiv:2110.14878 [hep-ex]].
		
	\bibitem{ParticleDataGroup:2020ssz}
	P.~A.~Zyla \textit{et al.} [Particle Data Group],
	PTEP \textbf{2020} (2020) no.8, 083C01.
	
	\bibitem{Awramik:2003rn}
	M.~Awramik, M.~Czakon, A.~Freitas and G.~Weiglein,
	Phys. Rev. D \textbf{69} (2004), 053006
	[arXiv:hep-ph/0311148 [hep-ph]].
			
	\bibitem{Peskin:1990zt}
	M.~E.~Peskin and T.~Takeuchi,
	Phys. Rev. Lett. \textbf{65} (1990), 964-967.
	
	\bibitem{Peskin:1991sw}
	M.~E.~Peskin and T.~Takeuchi,
	Phys. Rev. D \textbf{46} (1992), 381-409.


		

	\bibitem{Lu:2022bgw}
	C.~T.~Lu, L.~Wu, Y.~Wu and B.~Zhu,
	[arXiv:2204.03796 [hep-ph]].
		

\bibitem{Broggio:2014mna}
A.~Broggio, E.~J.~Chun, M.~Passera, K.~M.~Patel and S.~K.~Vempati,
JHEP \textbf{11} (2014), 058
[arXiv:1409.3199 [hep-ph]].

	\bibitem{Fan:2022dck}
	Y.~Z.~Fan, T.~P.~Tang, Y.~L.~S.~Tsai and L.~Wu,
	[arXiv:2204.03693 [hep-ph]].
	
	\bibitem{Song:2022xts}
	H.~Song, W.~Su and M.~Zhang,
	[arXiv:2204.05085 [hep-ph]].
		
	\bibitem{Bahl:2022xzi}
	H.~Bahl, J.~Braathen and G.~Weiglein,
	[arXiv:2204.05269 [hep-ph]].
	
	\bibitem{Babu:2022pdn}
	K.~S.~Babu, S.~Jana and V.~P.~K.,
	[arXiv:2204.05303 [hep-ph]].
	
	\bibitem{Biekotter:2022abc}
	T.~Biek\"otter, S.~Heinemeyer and G.~Weiglein,
	[arXiv:2204.05975 [hep-ph]].
	
	\bibitem{Han:2022juu}
	X.~F.~Han, F.~Wang, L.~Wang, J.~M.~Yang and Y.~Zhang,
	[arXiv:2204.06505 [hep-ph]].
	
	\bibitem{Heo:2022dey}
	Y.~Heo, D.~W.~Jung and J.~S.~Lee,
	[arXiv:2204.05728 [hep-ph]].
	
	\bibitem{Ahn:2022xeq}
	Y.~H.~Ahn, S.~K.~Kang and R.~Ramos,
	[arXiv:2204.06485 [hep-ph]].
	
	\bibitem{Benbrik:2022dja}
R.~Benbrik, M.~Boukidi and B.~Manaut,
[arXiv:2204.11755 [hep-ph]].
	
	\bibitem{Arcadi:2022dmt}
	G.~Arcadi and A.~Djouadi,
	[arXiv:2204.08406 [hep-ph]].
	
	\bibitem{Ghorbani:2022vtv}
	K.~Ghorbani and P.~Ghorbani,
	[arXiv:2204.09001 [hep-ph]].
	
	\bibitem{Lee:2022gyf}
	S.~Lee, K.~Cheung, J.~Kim, C.~T.~Lu and J.~Song,
	[arXiv:2204.10338 [hep-ph]].
	
	\bibitem{Du:2022brr}
	X.~K.~Du, Z.~Li, F.~Wang and Y.~K.~Zhang,
	[arXiv:2204.05760 [hep-ph]].
	
	\bibitem{Ghoshal:2022vzo}
	A.~Ghoshal, N.~Okada, S.~Okada, D.~Raut, Q.~Shafi and A.~Thapa,
	[arXiv:2204.07138 [hep-ph]].
	
	\bibitem{Kanemura:2022ahw}
	S.~Kanemura and K.~Yagyu,
	[arXiv:2204.07511 [hep-ph]].
	
	\bibitem{Addazi:2022fbj}
	A.~Addazi, A.~Marciano, A.~P.~Morais, R.~Pasechnik and H.~Yang,
	[arXiv:2204.10315 [hep-ph]].
	
	
    \bibitem{Yang:2022gvz}
    J.~M.~Yang and Y.~Zhang,
    [arXiv:2204.04202 [hep-ph]].

	\bibitem{Strumia:2022qkt}
    A.~Strumia,
    [arXiv:2204.04191 [hep-ph]].
    
    
    \bibitem{Sakurai:2022hwh}
    K.~Sakurai, F.~Takahashi and W.~Yin,
    [arXiv:2204.04770 [hep-ph]].
    
    \bibitem{Liu:2022jdq}
    X.~Liu, S.~Y.~Guo, B.~Zhu and Y.~Li,
    [arXiv:2204.04834 [hep-ph]].
    
    \bibitem{DiLuzio:2022xns}
    L.~Di Luzio, R.~Gr\"ober and P.~Paradisi,
    [arXiv:2204.05284 [hep-ph]].
		
	\bibitem{Asadi:2022xiy}
	P.~Asadi, C.~Cesarotti, K.~Fraser, S.~Homiller and A.~Parikh,
	[arXiv:2204.05283 [hep-ph]].
		
	\bibitem{Heckman:2022the}
	J.~J.~Heckman,
	[arXiv:2204.05302 [hep-ph]].
	
	\bibitem{Bagnaschi:2022whn}
	E.~Bagnaschi, J.~Ellis, M.~Madigan, K.~Mimasu, V.~Sanz and T.~You,
	[arXiv:2204.05260 [hep-ph]].
		
	\bibitem{Paul:2022dds}
	A.~Paul and M.~Valli,
	[arXiv:2204.05267 [hep-ph]].
	
	\bibitem{Balkin:2022glu}
	R.~Balkin, E.~Madge, T.~Menzo, G.~Perez, Y.~Soreq and J.~Zupan,
	[arXiv:2204.05992 [hep-ph]].
	
	
	\bibitem{Endo:2022kiw}
	M.~Endo and S.~Mishima,
	[arXiv:2204.05965 [hep-ph]].
			
	\bibitem{Zheng:2022irz}
	M.~D.~Zheng, F.~Z.~Chen and H.~H.~Zhang,
	[arXiv:2204.06541 [hep-ph]].
	
	
	
	\bibitem{Carpenter:2022oyg}
	L.~M.~Carpenter, T.~Murphy and M.~J.~Smylie,
	[arXiv:2204.08546 [hep-ph]].
	
	\bibitem{Popov:2022ldh}
	O.~Popov and R.~Srivastava,
	[arXiv:2204.08568 [hep-ph]].
	
	\bibitem{Chowdhury:2022moc}
	T.~A.~Chowdhury, J.~Heeck, S.~Saad and A.~Thapa,
	[arXiv:2204.08390 [hep-ph]].
	
	\bibitem{Cirigliano:2022qdm}
	V.~Cirigliano, W.~Dekens, J.~de Vries, E.~Mereghetti and T.~Tong,
	[arXiv:2204.08440 [hep-ph]].
	
	\bibitem{Bhaskar:2022vgk}
	A.~Bhaskar, A.~A.~Madathil, T.~Mandal and S.~Mitra,
	[arXiv:2204.09031 [hep-ph]].
	
	\bibitem{Baek:2022agi}
	S.~Baek,
	[arXiv:2204.09585 [hep-ph]].
	
	\bibitem{Cao:2022mif}
	J.~Cao, L.~Meng, L.~Shang, S.~Wang and B.~Yang,
	[arXiv:2204.09477 [hep-ph]].
    \bibitem{Kawamura:2022uft}
    J.~Kawamura, S.~Okawa and Y.~Omura,
    [arXiv:2204.07022 [hep-ph]].
    
    \bibitem{Nagao:2022oin}
    K.~I.~Nagao, T.~Nomura and H.~Okada,
    [arXiv:2204.07411 [hep-ph]].
    
    \bibitem{Zhang:2022nnh}
    K.~Y.~Zhang and W.~Z.~Feng,
    [arXiv:2204.08067 [hep-ph]].
    
	\bibitem{Borah:2022zim}
	D.~Borah, S.~Mahapatra and N.~Sahu,
	[arXiv:2204.09671 [hep-ph]].
	
	\bibitem{Cheng:2022aau}
	Y.~Cheng, X.~G.~He, F.~Huang, J.~Sun and Z.~P.~Xing,
	[arXiv:2204.10156 [hep-ph]].
	
	\bibitem{Batra:2022org}
	A.~Batra, S.~K.~A, S.~Mandal and R.~Srivastava,
	[arXiv:2204.09376 [hep-ph]].
		
	
	\bibitem{Branco:2011iw}
	G.~C.~Branco, P.~M.~Ferreira, L.~Lavoura, M.~N.~Rebelo, M.~Sher and J.~P.~Silva,
	Phys. Rept. \textbf{516} (2012), 1-102
	[arXiv:1106.0034 [hep-ph]].
	
	\bibitem{Glashow:1976nt}
	S.~L.~Glashow and S.~Weinberg,
	Phys. Rev. D \textbf{15} (1977), 1958
		
	\bibitem{Eriksson:2009ws}
	D.~Eriksson, J.~Rathsman and O.~Stal,
	Comput. Phys. Commun. \textbf{181} (2010), 189-205
	[arXiv:0902.0851 [hep-ph]].
	
	\bibitem{ATLAS:2015yey}
	G.~Aad \textit{et al.} [ATLAS and CMS],
	Phys. Rev. Lett. \textbf{114} (2015), 191803
	[arXiv:1503.07589 [hep-ex]].
	
	\bibitem{Haller:2018nnx}
	J.~Haller, A.~Hoecker, R.~Kogler, K.~M\"onig, T.~Peiffer and J.~Stelzer,
	Eur. Phys. J. C \textbf{78} (2018) no.8, 675
	[arXiv:1803.01853 [hep-ph]].
	
	\bibitem{Bechtle:2020pkv}
	P.~Bechtle, D.~Dercks, S.~Heinemeyer, T.~Klingl, T.~Stefaniak, G.~Weiglein and J.~Wittbrodt,
	Eur. Phys. J. C \textbf{80} (2020) no.12, 1211
	[arXiv:2006.06007 [hep-ph]].
	
	\bibitem{Bechtle:2020uwn}
	P.~Bechtle, S.~Heinemeyer, T.~Klingl, T.~Stefaniak, G.~Weiglein and J.~Wittbrodt,
	Eur. Phys. J. C \textbf{81} (2021) no.2, 145
	[arXiv:2012.09197 [hep-ph]].
	
	\bibitem{Mahmoudi:2008tp}
	F.~Mahmoudi,
	Comput. Phys. Commun. \textbf{180} (2009), 1579-1613
	[arXiv:0808.3144 [hep-ph]].
	
	\bibitem{ALEPH:2005ab}
	S.~Schael \textit{et al.} [ALEPH, DELPHI, L3, OPAL, SLD, LEP Electroweak Working Group, SLD Electroweak Group and SLD Heavy Flavour Group],
	Phys. Rept. \textbf{427} (2006), 257-454
	[arXiv:hep-ex/0509008 [hep-ex]].
	
	\bibitem{deBlas:2022hdk}
	J.~de Blas, M.~Pierini, L.~Reina and L.~Silvestrini,
	[arXiv:2204.04204 [hep-ph]].
	
	\bibitem{Arhrib:2016wpw}
	A.~Arhrib, R.~Benbrik and S.~Moretti,
	Eur. Phys. J. C \textbf{77} (2017) no.9, 621
	[arXiv:1607.02402 [hep-ph]].
	
	\bibitem{Bahl:2021str}
	H.~Bahl, T.~Stefaniak and J.~Wittbrodt,
	JHEP \textbf{06} (2021), 183
	[arXiv:2103.07484 [hep-ph]].
	
	\bibitem{Arhrib:2021xmc}
	A.~Arhrib, R.~Benbrik, M.~Krab, B.~Manaut, S.~Moretti, Y.~Wang and Q.~S.~Yan,
	JHEP \textbf{10} (2021), 073
	[arXiv:2106.13656 [hep-ph]].
	
	\bibitem{Wang:2021pxc}
	Y.~Wang, A.~Arhrib, R.~Benbrik, M.~Krab, B.~Manaut, S.~Moretti and Q.~S.~Yan,
	JHEP \textbf{12} (2021), 021
	[arXiv:2107.01451 [hep-ph]].
	
	\bibitem{Arhrib:2021yqf}
	A.~Arhrib, R.~Benbrik, M.~Krab, B.~Manaut, S.~Moretti, Y.~Wang and Q.~S.~Yan,
	Symmetry \textbf{13} (2021) no.12, 2319
	[arXiv:2110.04823 [hep-ph]].
	
\end{thebibliography}
\end{document}